\documentclass[pdflatex,sn-mathphys]{sn-jnl}

\usepackage{bm}
\usepackage{amsthm,amsmath,amssymb}
\usepackage{mathrsfs}

\jyear{2023}%

\theoremstyle{thmstyleone}%
\theoremstyle{thmstyletwo}%

\theoremstyle{thmstylethree}%

\begin{document}

\title[A unified framework for drug-target prediction]{DTIAM: A unified framework for predicting drug-target interactions, binding affinities and activation/inhibition mechanisms}

\author[1]{\fnm{Zhangli} \sur{Lu}}\email{luzhangli@csu.edu.cn}
\author[1]{\fnm{Chuqi} \sur{Lei}}\email{leichuqi@csu.edu.cn}
\author[1]{\fnm{Kaili} \sur{Wang}}\email{kailiwang@csu.edu.cn}
\author[1]{\fnm{Libo} \sur{Qin}}\email{lbqin@csu.edu.cn}
\author[2]{\fnm{Jing} \sur{Tang}}\email{jing.tang@helsinki.fi}
\author*[1]{\fnm{Min} \sur{Li}}\email{limin@mail.csu.edu.cn}

\affil*[1]{\orgdiv{School of Computer Science and Engineering}, \orgname{Central South University}, \orgaddress{\city{Changsha}, \postcode{410083}, \country{China}}}
\affil[2]{\orgdiv{Research Program in Systems Oncology, Faculty of Medicine}, \orgname{University of Helsinki}, \orgaddress{\city{Helsinki}, \postcode{FI00014}, \country{Finland}}}

\abstract{Accurate and robust prediction of drug-target interactions (DTIs) plays a vital role in drug discovery. Despite extensive efforts have been invested in predicting novel DTIs, existing approaches still suffer from insufficient labeled data and cold start problems. More importantly, there is currently a lack of studies focusing on elucidating the mechanism of action (MoA) between drugs and targets. Distinguishing the activation and inhibition mechanisms is critical and challenging in drug development. Here, we introduce a unified framework called DTIAM, which aims to predict interactions, binding affinities, and activation/inhibition mechanisms between drugs and targets. DTIAM learns drug and target representations from large amounts of label-free data through self-supervised pre-training, which accurately extracts the substructure and contextual information of drugs and targets, and thus benefits the downstream prediction based on these representations. DTIAM achieves substantial performance improvement over other state-of-the-art methods in all tasks, particularly in the cold start scenario. Moreover, independent validation demonstrates the strong generalization ability of DTIAM. All these results suggested that DTIAM can provide a practically useful tool for predicting novel DTIs and further distinguishing the MoA of candidate drugs. DTIAM, for the first time, provides a unified framework for accurate and robust prediction of drug-target interactions, binding affinities, and activation/inhibition mechanisms.}


\maketitle

\section{Introduction}\label{sec1}

Accurately predicting drug-target interactions (DTIs) is an essential step in drug discovery and development \cite{chen2016drug,wen2017deep}. The biochemical experimental method for identifying
new DTIs on a large scale is still expensive and time-consuming \cite{dimasi2003price,paul2010improve,pushpakom2019drug}, despite the wide application of various experimental assays in drug discovery. Various computational methods have been applied to drug discovery and successfully predict novel DTIs, and they can substantially reduce development time and costs \cite{zhang2022deepmgt,shin2019self,ye2021unified}. Current computational methods mainly focus on the binary prediction of DTI or the regression prediction of drug-target binding affinity (DTA). 

In binary classification-based DTI prediction studies, the goal is to predict whether there is an interaction between the drug and the target or not. Generally, the approaches for in silico DTI prediction can be divided into two major categories: structure-based approaches and structure-free approaches. Structure determination of compound-protein complexes can provide insights into the mode of action and thus significantly facilitate lead compound selection and optimization in the target-based drug discovery \cite{batool2019structure,yang2021g}. There are many structure-based approaches, such as molecular docking \cite{saikia2019molecular}, molecular dynamics simulations \cite{salo2020molecular}, pharmacophore modeling \cite{schaller2020next} and GOLD \cite{pagadala2017software}, which are widely applied in virtual screening of drugs binding with proteins. However, these methods generally fail to predict binding affinities when the three-dimensional (3D) structure of the target protein is unknown, and require tremendous computational resources. To overcome the current limitations of the structure-based methods, various structure-free models have been developed for DTI prediction \cite{yamanishi2008prediction,cichonska2017computational,mei2013drug,olayan2018ddr}. An example is the network-based inference (NBI) methods that construct reliable networks from several data resources (e.g., chemical, genomics, proteomics, and pharmacology) and exploit the topological and structural information in the networks for potential association prediction \cite{cheng2012prediction,fakhraei2014network,wu2018network,lu2019hnedti}. For instance, Luo et al. \cite{luo2017network} develop a computational pipeline, called DTINet, to predict novel DTIs from a heterogeneous network constructed by integrating diverse drug-related information. Another promising approach for predicting DTIs is the machine learning-based methods that mainly consist of two steps: feature extraction and DTI prediction \cite{bagherian2021machine,ding2014similarity,chen2018machine,d2020machine}. This type of approach fully exploits the latent features from input data of known drug compounds and target proteins to predict their interactions \cite{liu2015improving,shi2019predicting}. While these methods can successfully predict the interactions between each pair of drugs and targets, they fail to infer the strength of the interaction between the drug–target pairs.

In order to further predict the putative strengths of the interactions, various regression-based models have been proposed to infer the binding affinities between drugs and targets \cite{pahikkala2015toward,nguyen2021graphdta,jiang2020drug}. Binding affinity reflects how tightly the drug binds to a particular target, which is quantified by measures such as inhibition constant (Ki), dissociation constant (Kd), and the half-maximal inhibitory concentration (IC50). The DTA prediction approaches focus on affinity scoring, which is frequently used after virtual screening and docking campaigns. Recently, deep learning methods have emerged as a successful alternative to scoring functions, employing various deep neural network architectures such as convolutional neural network (CNN) and recurrent neural network (RNN). These methods fully extract contextual features and learn the representations of drugs and targets from the input raw data for DTA prediction. For example, DeepDTA \cite{ozturk2018deepdta} proposed by {\"O}zt{\"u}rk et al. used CNN to learn representations from the simplified molecular-input line-entry system (SMILES) strings of compounds and amino acid sequences of proteins, and fed into fully connected layers to predict their affinities. Karimi et al. \cite{karimi2019deepaffinity} presented a semi-supervised deep learning model, named DeepAffinity, which unifies RNN and CNN to jointly encode molecular and protein representations and predict affinities. Although these methods can successfully predict the binding affinity between drugs and targets, their interpretability remains limited. The attention mechanism has therefore been applied to increase the interpretability of the model by assigning greater weights to the ‘‘important’’ features \cite{vaswani2017attention,gao2018interpretable,li2022bacpi}. As an example, Li et al. \cite{li2020monn} developed a multi-objective neural network called MONN, which uses non-covalent interactions as additional supervision information to guide the model to capture the key binding sites.

While much effort has been devoted to predicting DTI and DTA, there are still several limitations in the previous studies. First, most existing methods heavily depend on the scale of the high-quality labeled data. Only large-scale labeled data can help models achieve great performance. Unfortunately, existing labeled data is insufficient, and data labeling is expensive and time-consuming. In addition, these methods often exhibit limited generalization when new drugs or targets are identified for a complicated disease, which is similar to the cold start problem in recommendation systems. More importantly, recent approaches fail to elucidate the mechanism of action (MoA) of the compound. The MoA refers to how a drug works on its target to produce the desired effects, which involve two major roles: activation and inhibition mechanisms. Distinguishing the activation and inhibition MoA between drugs and targets is critical and challenging in the drug discovery and development process, as well as their clinical applications \cite{zhang2023drugai}. It helps pharmaceutical scientists identify potential drug interactions and adverse effects, and develop safe and effective treatments for diseases \cite{schenone2013target,gibbs2000mechanism}. For example, drugs that activate dopamine receptors can treat Parkinson's disease, while drugs that inhibit dopamine receptors can treat psychosis \cite{sawada2018predicting}. 

In this study, we developed DTIAM, a unified framework for predicting DTI, DTA, and MoA. DTIAM learns drug and target representations from a large amount of unlabeled data via multi-task self-supervised pre-training, which requires only the molecular graph of drug compounds and primary sequences of target proteins as input. It accurately extracts the substructure and contextual information from massive compound and protein data during pre-training, which improves generalization performance and provides benefits for downstream tasks. In comprehensive comparison tests across different types of tasks and under three common and realistic experiment settings (i.e., warm start, drug cold start, and target cold start), DTIAM outperformed other baseline methods in all tasks, especially in the cold start scenario. Furthermore, independent validation on EFGR, CDK 4/6, and 10 specific targets indicated that DTIAM can provide a practically useful tool for predicting novel DTIs and further distinguishing the action mechanisms of candidate drugs. In addition, the robustness experiments demonstrated that the representations learned by the pre-training models transfer well to downstream tasks, even with limited labeled data for training. All these results suggested that DTIAM can provide accurate representations for effectively predicting candidate drug molecules or target proteins, and thus greatly facilitate the drug discovery process.

\section{Results}\label{sec2}

\subsection{Overview of DTIAM}

\begin{figure}[!tpb]%
\centering
\includegraphics[width=1.0\textwidth]{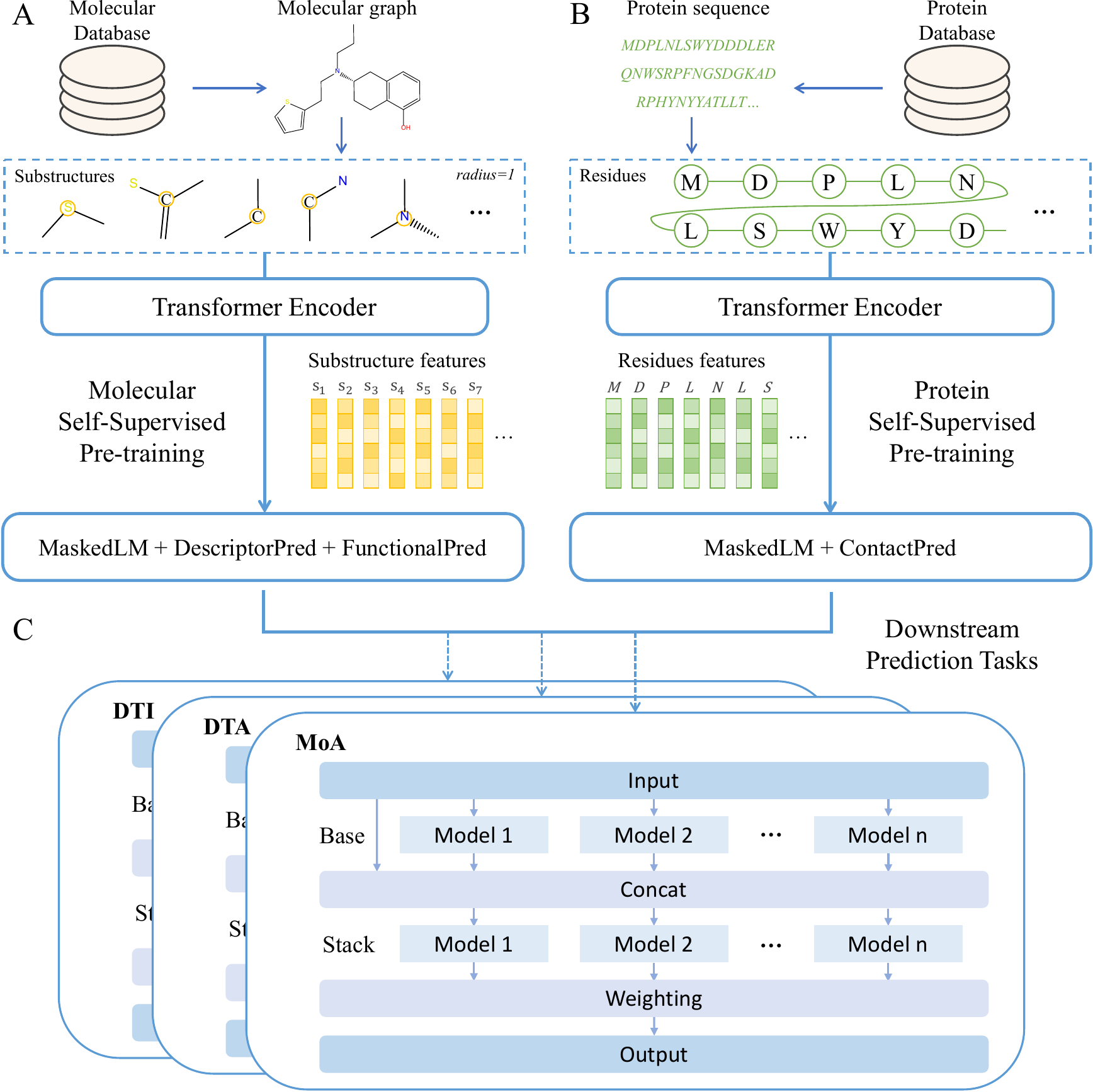}
\caption{\textbf{The architecture overview of DTIAM.} The framework mainly consists of three modules. (A) The drug molecular pre-training module. The module segments the molecular graph into several substructures and learns its representation through three self-supervised models from massive amounts of label-free data. (B) The target protein pre-training module. The module uses Transformer attention maps to learn the representations and contacts of proteins based on unsupervised language modeling from large amounts of protein sequence data. (C) The downstream drug-target prediction module. The module incorporates drug and target representation and predicts DTI, DTA, and MoA via an automated machine learning model.}\label{flowchart}
\end{figure}

Our proposed DTIAM is a general framework used for predicting DTI, DTA, and MoA based on self-supervised learning. The overall architecture of DTIAM is illustrated in Fig. \ref{flowchart}. DTIAM is not an end-to-end neural network model, which consists of three modules: (1) a drug molecular pre-training module based on multi-task self-supervised learning for extracting the features of both individual substructures and the whole compound from massive amounts of the molecular graph (Fig. \ref{flowchart}A), (2) a target protein pre-training module based on Transformer attention maps for extracting the features of individual residues directly from protein sequences (Fig. \ref{flowchart}B), and (3) a unified drug-target prediction module for predicting DTI, DTA, and MoA between the given pair of drug and target, using the previously learned features of drug molecular and target protein (Fig. \ref{flowchart}C). 

The drug molecule pre-training module takes the molecular graph as input, which is then segmented into several substructures. The module then learns the representation of the drug molecule based on multiple self-supervised models. Specifically, for a drug molecule with $n$ substructures, their representations are defined as a $n \times d$ embedding matrix, in which each substructure is embedded into a $d$-dimensional vector. These embeddings are fed into a Transformer encoder for feature extraction and learned through three self-supervised tasks: Masked Language Modeling, Molecular Descriptor Prediction, and Molecular Functional Group Prediction. The drug molecule pre-training module leverages the power of attention mechanism and self-supervised learning from vast amounts of unlabeled data to effectively extract contextual information and implicit features between molecular substructures. This process enables the module to learn meaningful representations of drug molecules without relying on explicit labels. By using the attention mechanism, the module can prioritize relevant substructures and relationships between them during training, leading to more effective feature extraction. Similarly, the target protein pre-training module uses Transformer attention maps to learn the representations and contacts of proteins based on unsupervised language modeling from large amounts of protein sequence data. The basic idea of the drug-target prediction module is to integrate information from both drugs and targets to improve the prediction of DTI, DTA, and MoA. The module combines representations of compounds and proteins to capture their complex interactions and uses various machine learning models, such as neural networks, to learn their relationship and properties for accurate and reliable predictions. These models are integrated within an automated machine learning framework that utilizes multi-layer stacking and bagging techniques. Details about each module of DTIAM and the training process can be found in section \ref{sec4} (Methods).

\begin{figure}[!tpb]
\centering
\includegraphics[width=1.0\textwidth]{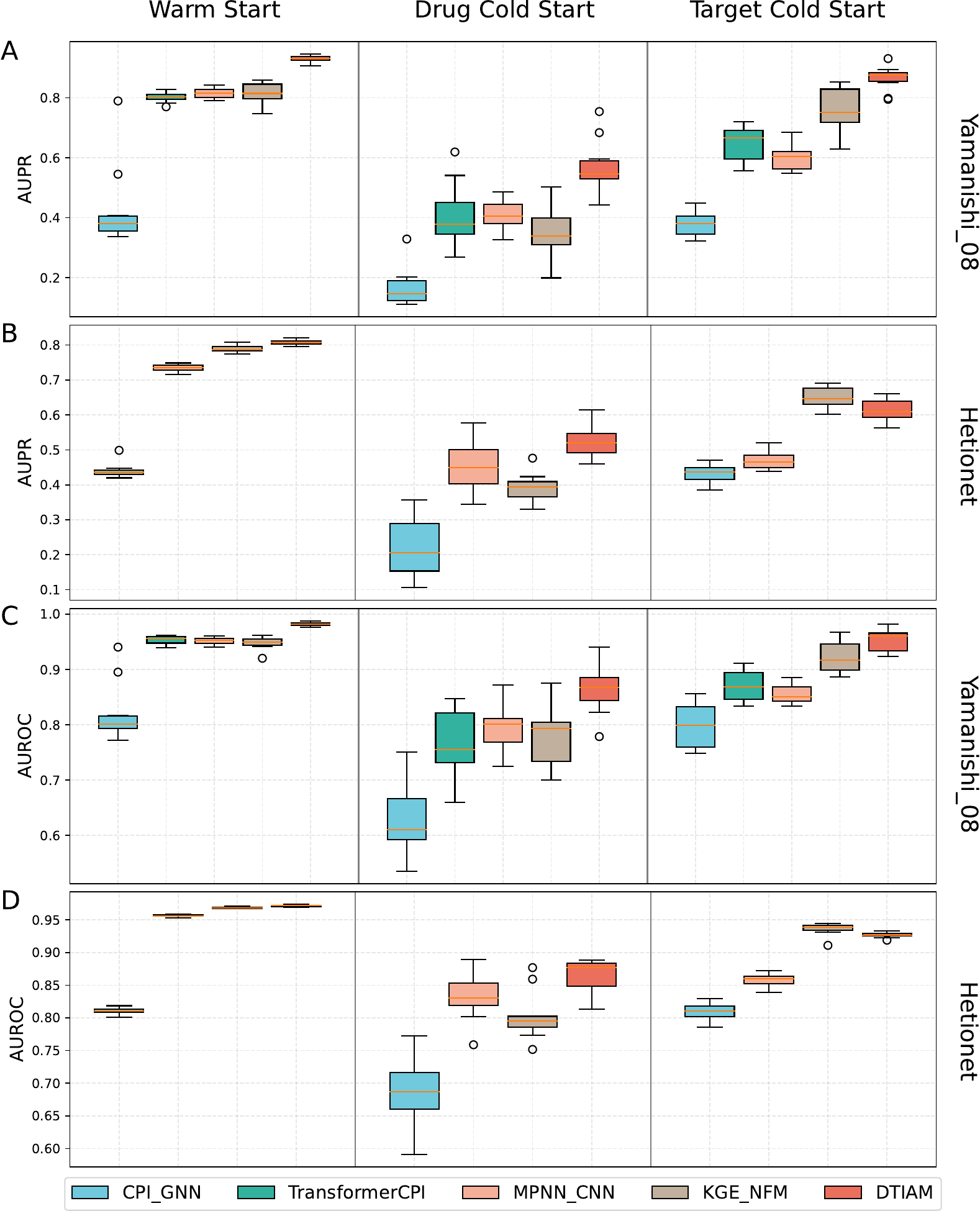}
\caption{\textbf{Performance evaluation on the DTI prediction task.} The performances of DTIAM and baseline models were evaluated in terms of AUPR and AUROC on the Yamanishi\_08’s and Hetionet datasets under three experiment settings. (A-B) AUPR of different prediction models on the Yamanishi\_08’s and Hetionet datasets under three experiment settings. (C-D) AUROC of different prediction models on the Yamanishi\_08’s and Hetionet datasets under three experiment settings. All results were obtained by 10-fold cross-validation. The ratio between the positive and negative samples is 1:10.}\label{res_dti}
\end{figure}

\subsection{Performance of DTIAM on the DTI prediction task}

In the DTI prediction task, the goal is to predict whether a given drug-target pair interacts with each other, which is a binary classification problem. We compared DTIAM with four baseline methods, including CPI\_GNN \cite{tsubaki2019compound}, TransformerCPI \cite{chen2020transformercpi}, MPNN\_CNN \cite{gilmer2017neural}, and KGE\_NFM \cite{ye2021unified}, on the Yamanishi\_08’s and Hetionet benchmark datasets under three experiment settings (Fig. \ref{res_dti}, Supplementary Materials Table 3). The training data and test data are split via 10-fold cross-validation as the previous studies \cite{luo2017network,ye2021unified,mohamed2020discovering,thafar2020dtigems}, and the ratio between the positive and negative samples is 1:10.

First, on the smaller Yamanishi\_08’s dataset, we observed that DTIAM achieved higher and more robust predictive performance under three different experiment settings, especially in the cold start settings. Specifically, in the scenario of the warm start, DTIAM (AUPR=0.931) significantly outperformed all the other baselines with a significant leading margin of 50\% in terms of AUPR when compared to CPI\_GNN (AUPR=0.431). While for the end-to-end methods, TransformerCPI (AUPR=0.816) and MPNN\_CNN (AUPR=0.802), and the network-based method KGE\_NFM (AUPR=0.817) achieved comparable predictive performance. These results indicate that the end-to-end methods and network-based methods require more labeled data, while DTIAM can partly overcome this limitation thanks to the knowledge learned in the pre-training stage. In the scenario of the cold start, we observed that the AUPR and AUROC values of all methods get reduced by different degrees, while DTIAM still achieves relatively high predictive performance, especially in the target cold start. This result highlights DTIAM’s potential capability to capture the latent features of compound substructures and protein subsequences from the large-scale unlabeled data, thus enables higher accuracy and more robust prediction even for unknown drugs or targets. 

On the other hand, in the larger Hetionet dataset, we observed that DTIAM achieved the better, the best, and the second best predictive performance in the warm start, the drug cold start, and the target cold start, respectively. Specifically, in the scenario of the warm start, the average AUPR score achieved by DTIAM (AUPR=0.808) was higher than other baseline methods. While for the network-based method, KGE\_NFM (AUPR=0.789) achieved comparable predictive performance due to the increased volume of available data. In the scenario of the drug cold start, DTIAM (AUPR=0.529) significantly outperformed CPI\_GNN (AUPR=0.219), MPNN\_CNN (AUPR=0.453), and KGE\_NFM (AUPR=0.391). This phenomenon demonstrates the powerful expressive and feature learning ability of the proposed drug pre-training model, which provides a huge advantage for DTIAM in the situation of unknown drug prediction. In the scenario of the target cold start, KGE\_NFM (AUPR=0.651) performed better than CPI\_GNN (AUPR=0.433), MPNN\_CNN (AUPR=0.470), and DTIAM (AUPR=0.614). This is mainly attributed to the sufficient target-related association information for the network-based method KGE\_NFM. While DTIAM and the end-to-end methods only take the compound SMILES and the protein sequences as input without extra association information. Additionally, we found a similar phenomenon on the Yamanishi\_08’s and Hetionet datasets that all methods achieved better predictive performance in the target cold start than the drug cold start. It seems possible that this finding is attributed to the volume of available data for targets, where both datasets have more targets than drugs. For example, there are 5,763 targets while only 1,384 drugs are in the Hetionet dataset (Supplementary Materials Table 1).

\begin{figure}[!tpb]%
\centering
\includegraphics[width=1.0\textwidth]{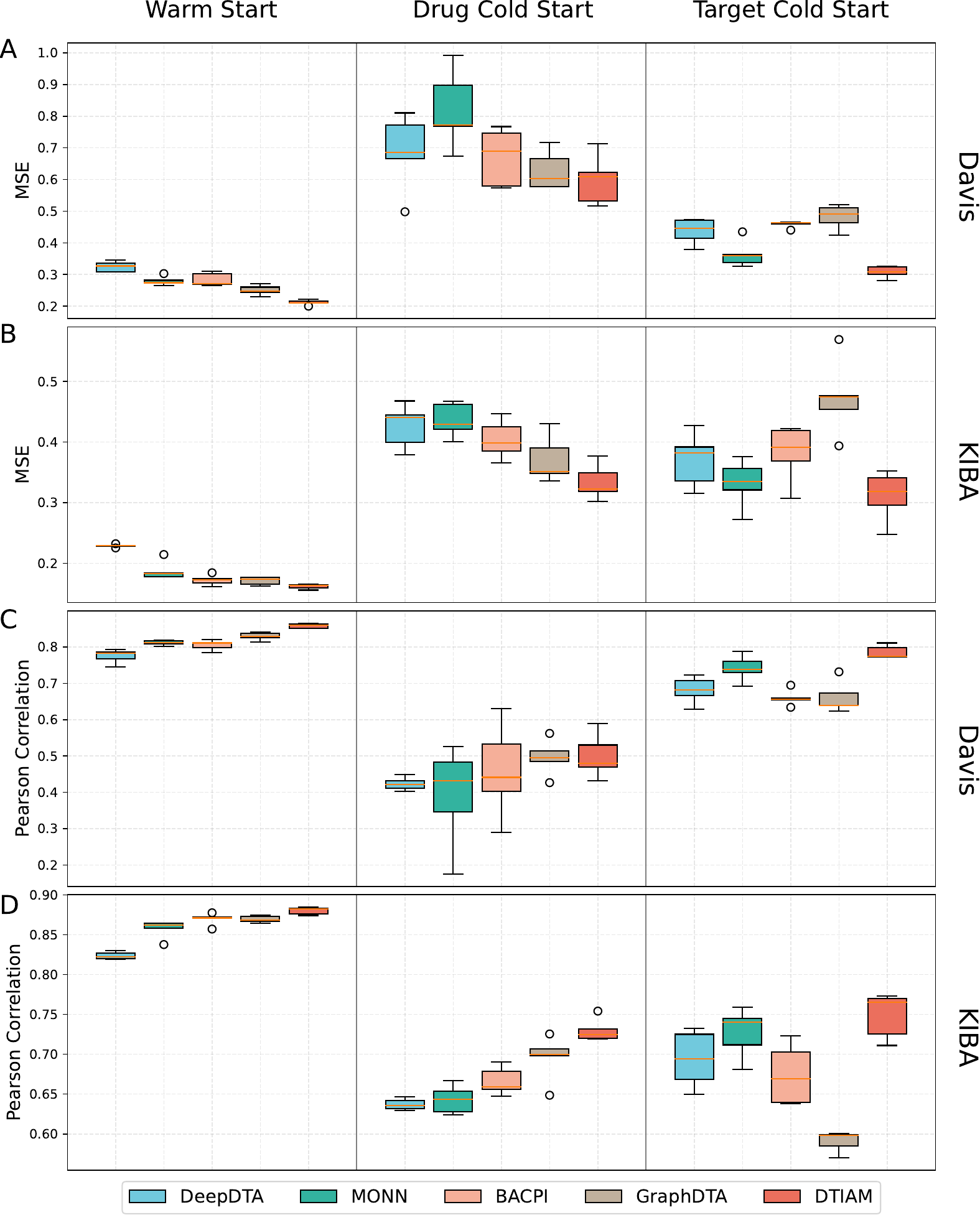}
\caption{\textbf{Performance evaluation on the DTA prediction task.} The performances of DTIAM and baseline models were evaluated in terms of MSE and Pearson correlation on the Davis and KIBA datasets under three experiment settings. (A-B) MSE of different prediction models on the Davis and KIBA datasets under three experiment settings. (C-D) Pearson correlation of different prediction models on the Davis and KIBA datasets under three experiment settings. All results were obtained by 5-fold cross-validation.}\label{res_dta}
\end{figure}

\subsection{Performance of DTIAM on the DTA prediction task}

The goal of the DTA prediction task is to predict the binding affinity between a given pair of drug targets, which is a regression task. And four baseline models were used in the performance comparison, including DeepDTA \cite{ozturk2018deepdta}, MONN \cite{li2020monn}, BACPI \cite{li2022bacpi}, and GraphDTA \cite{nguyen2021graphdta}. We evaluated our model and all the baseline methods on two benchmark datasets, the Kinase dataset Davis and KIBA dataset, under three experiment settings (Fig. \ref{res_dta}, Supplementary Materials Table 4). For each experiment setting, we use 5-fold cross-validation to evaluate the DTA prediction performance of DTIAM and baseline methods.

As can be seen from Fig. \ref{res_dta}, DTIAM achieved better predictive performance under all experimental settings on both datasets, especially in the cold start settings. For the scenario of the warm start, DTIAM and three graph-based methods, MONN, BACPI, and GraphDTA, achieved relatively high predictive performance on both datasets. While for the sequence-based method, DeepDTA did not perform as well due to the limitation of the model structure, which fails to extract accurate features from sequence information. In the scenario of the cold start, we observed a similar situation with the DTI prediction task in that the predictive performance gets reduced by different degrees for all methods. GraphDTA achieved relatively high predictive performance in the drug cold start setting, but do not perform as well in the target cold start setting. In contrast, MONN performed better in the target cold start setting than in the drug cold start setting. These results suggested that GraphDTA is more suitable for the binding affinity prediction of new drugs, while MONN is better for the situation of the cold start for targets. DeepDTA and BACPI behaved more stably in two cold start scenarios, which shows the robustness of the predictions. For the pre-training model, DTIAM performed the best in both the warm start setting and two cold start settings. All these comparative results supported the strong predictive power of DTIAM, which can successfully predict the binding affinities between drugs and targets, and has a strong generalization ability even for predictions on novel drugs or targets.

\begin{figure}[!tpb]%
\centering
\includegraphics[width=1.0\textwidth]{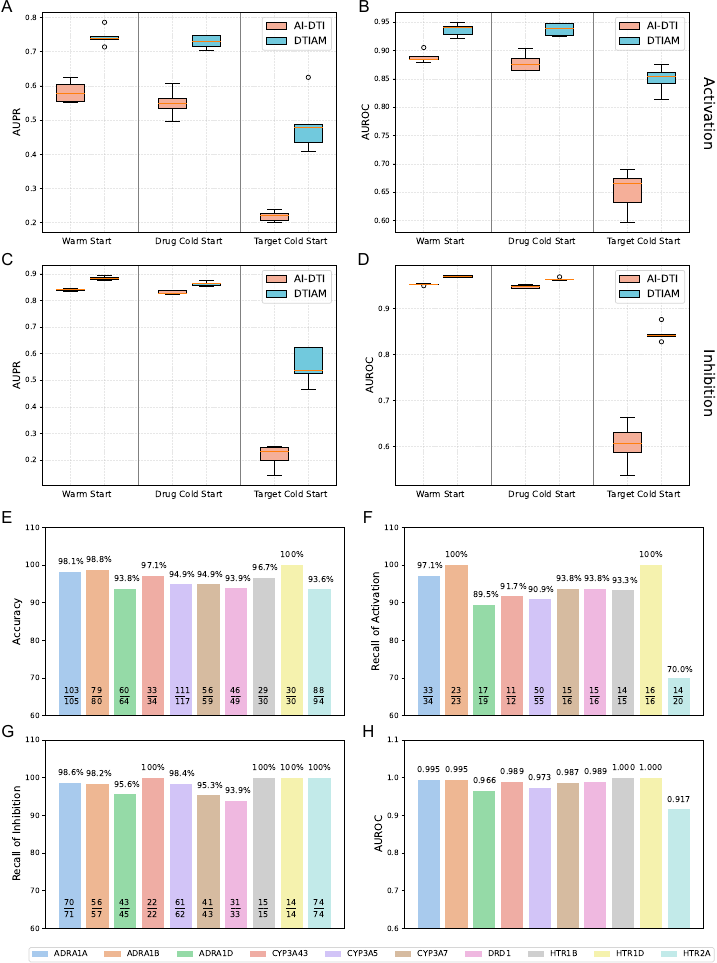}
\caption{\textbf{Performance evaluation on the MoA prediction task.} (A-D) Performance comparison of DTIAM with AI-DTI on the Activation and Inhibition datasets under three experiment settings in terms of AUPR and AUROC. All results were obtained by 5-fold cross-validation. The ratio between the positive and negative samples is 1:10. (E-H) Validating DTIAM on 10 specific targets. (E) Prediction accuracy for each target, with the score at the top of the bar and the ratio inside indicating the number of correctly predicted drugs out of the total number. (F-G) The recall of activation and inhibition, respectively, with the top value indicating recall and the ratio inside indicating the number of correctly predicted activators or inhibitors out of the total number. (H) AUROC score for each target.}\label{res_moa}
\end{figure}

\subsection{Performance of DTIAM on the MoA prediction task}

We first evaluated the prediction performance of DTIAM using various cross-validation settings, including three commonly used and more realistic experimental scenarios: the warm start setting, the drug cold start setting, and the target cold start setting. In this study, we approach MoA prediction as two distinct binary classification tasks. Specifically, we aim to predict whether a given drug-target pair exhibits activatory or inhibitory effects, based on the previous studies \cite{sawada2018predicting,lee2021predicting}. Thus, our two classification problems are predicting the presence of activation and inhibition for a given drug-target pair. The evaluations were conducted using two distinct types of MoA (i.e., activation and inhibition), and each type of MoA has a corresponding dataset collected from the Therapeutic Target Database. We compared DTIAM with AI-DTI \cite{lee2021predicting} using a 5-fold cross-validation on these two datasets under three experiment settings (Fig. \ref{res_moa}A-D, Supplementary Materials Table 2). 

As shown in Fig. \ref{res_moa}A-D, DTIAM significantly outperformed AI-DTI in all three experiment settings, on both Activation and Inhibition datasets. Specifically, on the Activation dataset, DTIAM outperformed AI-DTI with 16.1\%, 17.9\%, and 26.8\% improvement in terms of AUPR in the scenario of the warm start, the drug cold start, and the target cold start, respectively. These results indicate that the representations learned by self-supervised pre-training transfer well to downstream tasks without requiring large amounts of labeled training data. In particular, with a larger size of labeled drug-target pairs on the Inhibition dataset, the evaluation performance of DTIAM and AI-DTI increase greatly compared to that on the Activation dataset. And DTIAM performed slightly better than AI-DTI in the scenario of the warm start and the drug cold start. While DTIAM outperformed AI-DTI with a significant leading margin of 34\% in terms of AUPR when the experiment setting is the target cold start. This result demonstrated that the target protein pre-training model has the potential to learn patterns from large-scale protein amino acid sequences, and thus benefits the downstream prediction for DTIAM in the situation of the target cold start. In addition, we also observed an interesting phenomenon in the Activation and Inhibition datasets that both methods achieved better predictive performance in the drug cold start than the target cold start. This result may be explained by the fact that there are more drugs than targets in the two MoA datasets. This finding manifests the influence of the size of the predicted object in the scenario of the cold start, and a larger number of the predicted object enable better prediction performance.

\subsection{Validation of the activation/inhibition predictions}

Distinguishing the activation/inhibition mechanism between a drug and its target is of great biological significance because it can determine the type of biological response produced by the drug. Take the alpha-1A adrenergic receptor (ADRA1A) for example, drugs such as metaraminol activate ADRA1A for the treatment of hypotension \cite{kee2003hemodynamic}, whereas drugs inhibit ADRA1A used for benign prostatic hyperplasia, hypertension, schizophrenia, and bipolar disorder \cite{yono2004doxazosin,corena2015comparative}. To demonstrate the reliability of DTIAM in distinguishing activation and inhibition interactions, we combined all activating and inhibiting DTIs to train DTIAM, which is applied to predict the activation and inhibition relationships for 10 specific targets (including ADRA1A, ADRA1B, ADRA1D, CYP3A43, CYP3A5, CYP3A7, DRD1, HTR1B, HTR1D, and HTR2A). These targets can be categorized into four distinct subfamilies: alpha-1 adrenergic receptors (ADRA1), cytochrome P450 3A enzymes (CYP3A), dopamine receptors (DR), and 5-hydroxytryptamine receptors (HTR). Each of these subfamilies possesses a considerable repertoire of known agonists and antagonists, along with distinct mechanisms of action that align with diverse therapeutic indications. The exploration of the mechanisms of interaction between these subfamilies and drugs holds paramount importance in the realms of nervous system regulation, catalytic reactions, and beyond. We collect the activation/inhibition relationships for these 10 targets from DrugBank, and all of these relationships are independent of all the training data used for DTIAM. 

We list all prediction results in Supplemental Spreadsheet 1 and show the results in terms of accuracy, recall of activation, recall of inhibition, and AUROC in Fig. \ref{res_moa}E-H. We found that the prediction accuracy exceeded 93\% for all targets, including 100\% for HTR1D, and 9 of the targets had AUROC values above 0.96. These results suggest that DTIAM can accurately distinguish the activation and inhibition relationships between drugs and targets. In addition, we observed that the recall of activation was significantly lower than that of inhibition. This is because the samples of the dataset used in DTIAM are out of balance (far fewer samples for activation than for inhibition), which leads to a more biased prediction result of the model with inhibition. 

\begin{table}[!tpb]
\renewcommand{\arraystretch}{1.2}
\caption{Top-15 predicted candidate drugs for EGFR.}\label{case1}%
\begin{tabular}{ccccc}
\toprule
Rank & KEGG ID & Drug Name                       & Pred\_Score & Validation      \\ 
\midrule
1    & D01441       & Imatinib mesylate          & 0.9962      & BRENDA          \\ 
2    & D01977       & \textbf{Gefitinib}         & 0.9961      & DrugBank        \\ 
3    & D04023       & Erlotinib hydrochloride    & 0.9954      & KEGG,BRENDA     \\ 
4    & D04024       & \textbf{Lapatinib}         & 0.9941      & DrugBank        \\ 
5    & D03218       & Axitinib                   & 0.9937      & BRENDA          \\ 
6    & D04025       & Mubritinib                 & 0.9918      & KEGG            \\ 
7    & D03252       & Bosutinib                  & 0.9914      & KEGG            \\ 
8    & D03350       & Canertinib dihydrochloride & 0.9851      & BRENDA,DrugBank \\ 
9    & D09883       & \textbf{Dacomitinib}       & 0.9750      & DrugBank        \\ 
10   & D10766       & \textbf{Osimertinib}       & 0.9359      & DrugBank        \\ 
11   & D08950       & \textbf{Neratinib}         & 0.9308      & DrugBank        \\ 
12   & D09724       & \textbf{Afatinib}          & 0.9143      & DrugBank        \\ 
13   & D07907       & \textbf{Erlotinib}         & 0.9099      & DrugBank        \\ 
14   & D10866       & \textbf{Brigatinib}        & 0.8799      & DrugBank        \\ 
15   & D12001       & \textbf{Mobocertinib}      & 0.8106      & DrugBank        \\ 
\botrule
\end{tabular}
\footnotetext{\emph{Note}: The bolded drugs are the approved EGFR inhibitors collected from DrugBank, the others are the drugs in the Yamanishi\_08’s dataset. The column \emph{Pred\_Score} indicates the predicted probability of the drug candidate interacting with EGFR.}
\end{table}

\subsection{Validation of the top-ranked predictions}

To test the applicability for drug virtual screening, we tested whether DTIAM could identify the DTIs of candidate drugs for epidermal growth factor receptor (EGFR) and cyclin-dependent kinase 4/6 (CDK 4/6). EGFR is a transmembrane protein that is found at abnormally high levels in cancer cells, and its inhibitors are known for the treatment of cancers caused by EGFR up-regulation, such as non-small-cell lung cancer and pancreatic cancer. CDK is a type of enzyme that regulates the progression of cells through the cell cycle. CDK 4/6 inhibitors work by binding to and blocking the activity of CDK4 and CDK6 enzymes and are commonly used to treat breast cancer and other types of cancer that are driven by overactive CDK 4/6 activity. 

\begin{table}[!tpb]
\renewcommand{\arraystretch}{1.2}
\setlength\tabcolsep{2pt}
\caption{Top-15 predicted candidate drugs for CDK 4 and CDK 6.}\label{case2}%
\begin{tabular}{cccccc}
\toprule
Rank & KEGG ID & Drug Name                     & Pred\_Score1 & Pred\_Score2 & Validation   \\ 
\midrule
1    & D01441  & Imatinib mesylate             & 1.0000       & 1.0000       & CTD          \\ 
2    & D10883  & \textbf{Ribociclib}           & 0.9995       & 1.0000       & DrugBank     \\ 
3    & D02880  & Alvocidib hydrochloride       & 1.0000       & 1.0000       & KEGG         \\ 
4    & D01840  & Fasudil hydrochloride         & 0.9997       & 1.0000       & -            \\ 
5    & D10688  & \textbf{Abemaciclib}          & 0.9843       & 0.9999       & DrugBank     \\ 
6    & D09868  & Alvocidib                     & 0.9998       & 0.9999       & KEGG         \\ 
7    & D11130  & \textbf{Trilaciclib}          & 0.9347       & 0.9995       & DrugBank     \\ 
8    & D03350  & Canertinib dihydrochloride    & 0.9735       & 0.9994       & -            \\ 
9    & D04370  & Granisetron                   & 0.9920       & 0.9992       & -            \\ 
10   & D03736  & Doramapimod                   & 0.9974       & 0.9986       & -            \\ 
11   & D03218  & Axitinib                      & 0.9901       & 0.9986       & -            \\ 
12   & D03115  & Fasudil hydrochloride hydrate & 0.9964       & 0.9973       & -            \\ 
13   & D02217  & Raloxifene hydrochloride      & 0.9971       & 0.9942       & -            \\ 
14   & D10372  & \textbf{Palbociclib}          & 0.5924       & 0.9832       & DrugBank     \\ 
15   & D04025  & Mubritinib                    & 0.9755       & 0.9758       & -            \\ 
\botrule
\end{tabular}
\footnotetext{\emph{Note}: The bolded drugs are the approved CDK 4 and CDK 6 inhibitors collected from DrugBank, the others are the drugs in the Yamanishi\_08’s dataset. The column of \emph{Pred\_Score1} and \emph{Pred\_Score2} indicate the predicted probabilities of drug candidates interacting with CDK 4 and CDK 6, respectively. The drug candidates are ranked according to \emph{Pred\_Score2}.}
\end{table}

We used Yamanishi\_08’s dataset (removing DTIs containing EGFR) to train DTIAM and predict potential interactions between EGFR and all drugs in the dataset. We also predicted the interactions between EGFR and 13 approved drugs (Afatinib, Osimertinib, Gefitinib, Erlotinib, Lapatinib, Neratinib, Brigatinib, Dacomitinib, Mobocertinib, Vandetanib, Fostamatinib, Zanubrutinib, and Lidocaine) from DrugBank which are used as EGFR inhibitors. The predicted results of the top-15 drug candidates are listed in Table \ref{case1}. We found that 9 of the 13 EGFR inhibitors were successfully rediscovered in the top-15 drug candidates by our method, and 12 of 13 EGFR inhibitors ranked in the top 50 of 802 results (more details in Supplemental Spreadsheet 2). In addition, the other 6 drugs in the top-15 candidate list were all validated by external databases (e.g., KEGG \cite{kanehisa2006genomics}, BRENDA \cite{schomburg2004brenda}, and DrugBank \cite{wishart2008drugbank}). 

\begin{figure}[!tpb]%
\centering
\includegraphics[width=1.0\textwidth]{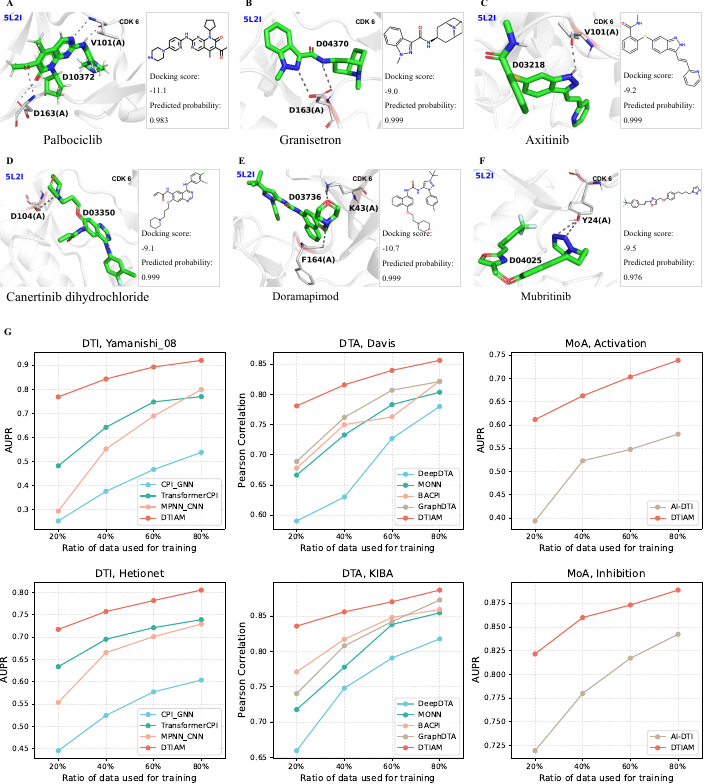}
\caption{\textbf{Docking results of the predicted interactions and performance comparison on different scales of labeled data.} (A-F) The docked poses for the predicted interactions between six candidate drugs (i.e., Palbociclib, Granisetron, Axitinib, Canertinib dihydrochloride, Doramapimod, and Mubritinib, where Palbociclib is the reference drug) and the CDK 6. (G) The performances of DTIAM and baseline models trained on different scales of labeled data (i.e., 20\%, 40\%, 60\%, and 80\%) in DTI, DTA, and MoA prediction tasks under the warm start settings.}\label{docking_robust}
\end{figure}

Similarly, the Yamanishi\_08’s dataset (removing DTIs containing CDK 4 and CDK 6) was used to train DTIAM, which is applied to predict the interactions between CDK 4/6 and all drugs in the dataset and 4 approved drugs (Ribociclib, Abemaciclib, Trilaciclib, and Palbociclib) from DrugBank which are used as inhibitors of CDK 4 and CDK 6. Table \ref{case2} shows the top-15 drug candidates that potentially interact with CDK 4/6, ranked by the prediction scores of CDK 6. We observed that 4 approved CDK 4/6 inhibitors are successfully predicted by DTIAM. Moreover, Imatinib mesylate, Alvocidib, and Alvocidib hydrochloride were validated to interact with CDK 4/6 by external databases (CTD \cite{davis2021comparative} and KEGG \cite{kanehisa2006genomics}). In addition, the docking studies showed that the five drugs (i.e., Granisetron, Axitinib, Canertinib dihydrochloride, Doramapimod, and Mubritinib) were able to dock to the CDK 6 (Fig. \ref{docking_robust}B-F). In particular, Granisetron interacted with residue D163(A) and Axitinib interacted with residue V101(A) when docked to CDK 6 (Fig. \ref{docking_robust}B, C), which were observed similar to Palbociclib (Fig. \ref{docking_robust}A), the highly selective CDK4/6 inhibitor. All these results indicated that DTIAM can be effectively applied for drug virtual screening and provide a powerful tool to speed up the process of drug development.

\subsection{Performance comparison of DTIAM with different scales of labeled data}

As mentioned above, DTIAM achieves excellent performance in downstream tasks even with a small amount of labeled data thanks to the pre-trained model. To test the effectiveness of our proposed pre-trained model, we compared DTIAM with the state-of-the-art baseline models with different scales of labeled data on the DTI, DTA, and MoA prediction tasks. We divided 20\%, 40\%, 60\%, and 80\% of the samples on six datasets of different tasks for training, and used the remaining samples for validation under the warm start setting.

As can be seen from Fig. \ref{docking_robust}G, with the increase of the number of training samples, the predictive performance of all methods improves by different degrees. DTIAM achieves the best performance under all data partitions and significantly outperforms other models, especially with less training data (20\% and 40\%). It is worth noting that DTIAM can outperform other models trained with 60\% or even 80\% of the samples using only 20\% of the samples for supervised training. All the results show that our proposed pre-training model can extract accurate features from massive unlabeled data and can be effectively applied to downstream prediction tasks, even with a small amount of unlabeled data.

\section{Discussion}\label{sec3}

Accurately predicting DTIs can provide a huge advantage for drug discovery and development. Most existing methods only focus on the DTI binary classification or the DTA regression prediction, neglecting the pharmaceutical MoA information. The MoA prediction can help in understanding modes of drug action and provide new insights into drug discovery. In this study, we developed a unified framework, called DTIAM, to predict DTI, DTA, and MoA by combining the drug and target pre-training models and AutoML techniques. The pre-training models extract the substructure and contextual information from massive unlabeled data via self-supervised learning and can be transferred to various prediction tasks including MoA, DTI, DTA, etc. The high extendibility and generalization ability of the pre-training models have been extensively validated on different types of prediction tasks. Comprehensive comparison tests showed that DTIAM achieved superior performance and significantly outperformed other state-of-the-art machine learning methods on different types of datasets under three cross-validation settings. Besides, the independent validation experiments on 10 specific targets demonstrate the ability of DTIAM to accurately discriminate the drug mechanisms of action. We also validated the applicability of DTIAM for drug virtual screening on EGFR and CDK 4/6 targets, the results showed that the top-15 predicted drug candidates were mostly validated by external databases and literature. All of these results demonstrate that DTIAM can be effectively used for a variety of drug target prediction tasks and provides a powerful tool for drug development.

DTIAM uses only molecular SMILES and protein sequences as input, and it effectively improves the performance of downstream prediction tasks by using massive amounts of label-free data for self-supervised pre-training. Previous studies have demonstrated that protein 3D structures are indeed useful in providing accurate target binding pockets as well as binding sites. Currently, various highly accurate protein structure prediction methods and molecular conformation generation algorithms have been proposed, such as AlphaFold2 \cite{jumper2021highly}, RoseTTAFold \cite{humphreys2021computed}, and GEODIFF \cite{xu2022geodiff}, which can help identify potential drug targets and facilitate the drug development process. In future work, we will integrate protein 3D structure and molecular conformation information to further improve prediction accuracy and model interpretability. 

\section{Methods}\label{sec4}

\subsection{The workflow of DTIAM}

DTIAM consists of three main components: (1) Self-supervised molecular representation Learning for drug pre-training; (2) Unsupervised protein representation Learning for target pre-training; (3) The representation integration and downstream drug target inference tasks via automated machine learning (AutoML).

\subsubsection{Self-supervised molecular representation learning}

We adopt the BERT-style \cite{devlin2018bert} method for drug pre-training and develop a molecular representation learning model called \textbf{BERMol}, which stands for \textbf{B}idirectional \textbf{E}ncoder \textbf{R}epresentations of \textbf{Mol}ecular. BERMol learns vector representations of molecular substructures from large-scale unlabeled data with the language model and domain-relevant auxiliary tasks. The proposed model is pre-trained on the GuacaMol dataset \cite{brown2019guacamol} which contains 1.6 million compounds collected from the ChEMBL \cite{gaulton2017chembl} database. To apply the language model to molecular, we define the substructures of molecules as ``words'' and molecules as ``sentences''. We then use the Morgan algorithm \cite{rogers2010extended} to extract all substructures of radius 1 for each molecule. After generating the corpus of compounds, the Transformer \cite{vaswani2017attention} architecture is applied to learn the low-rank representations for all substructures of molecules. Specifically, a molecule can be abstracted as a sentence $\bm{S} = ({x}_{1}, ..., {x}_{n})$, where ${x}_{i}$ is the $i$-th word and $n$ is the sentence length, each word in the sentence is then embedded into a $d$-dimensional vector space $\bm{Z} = (\vec{z}_{1}, ..., \vec{z}_{n})$, where $\vec{z}_{i} \in \mathbb{R}^{d}$ is the $d$-dimensional embedding of the $i$-th word. In the encoding step using the transformer architecture, we transform all embeddings into three matrices $(\bm{Q}, \bm{K}, \bm{V})$ representing queries, keys, and values respectively, and then compute the self-attention weights between words as follows:

\begin{equation}
{\rm Attention}(\bm{Q}, \bm{K}, \bm{V}) = {\rm softmax}(\bm{Q}\bm{K}^{T}/\sqrt{d})\bm{V}
\end{equation}
where $\sqrt{d}$ is the scaling factor used to smooth the gradient of the softmax function, and the output of the attention mechanism is a matrix representing the global relationship between different words. To integrate information from different representation subspaces, multi-head attention is performed with different linear projections, the final output matrix can be written as,

\begin{equation}
{\rm MultiHead}(\bm{Q}, \bm{K}, \bm{V}) = {\rm Concat}({\rm head}_{1}, ..., {\rm head}_{k})\bm{W}^{O}
\end{equation}

\begin{equation}
{\rm head}_{i} = {\rm Attention}(\bm{QW}_{i}^{Q}, \bm{KW}_{i}^{K}, \bm{VW}_{i}^{V})
\end{equation}
where $\bm{W}_{i}^{Q}, \bm{W}_{i}^{K}, \bm{W}_{i}^{V}$ are the projection matrices of $i$-th head. The complete encoder is a stack of multiple blocks combined with a multi-head self-attention mechanism and a fully connected feed-forward network. 

In order to learn flexible and high-quality molecular representations, we combine three self-supervised tasks for pre-training: (1) Masked Language Modeling (MLM); (2) Molecular Descriptor Prediction (MDP);  (3) Molecular Functional Group Prediction (MFGP). The MLM task was proposed by BERT, whereby the model randomly masks a portion of the tokens and is trained to predict the true identity of the masked tokens. In this task, the final representations of the masked tokes are fed into a neural network model for multiclassification prediction. The task is optimized using the cross-entropy loss as follows:

\begin{equation}
{\rm Loss}_{MLM} = - \frac{1}{N_{mask}}\sum_{i \in mask}\sum_{j=1}^{V}y_{ij}log(p_{ij})
\end{equation}
where $N_{mask}$ is the number of the masked tokens, $V$ is the size of the vocabulary, i.e. the size of the set of substructures, $y_{ij}$ is a one-hot vector representing the true distribution over the vocabulary for the $i$-th masked token, and $p_{ij}$ is the predicted probability of the $j$-th token in the vocabulary being the correct replacement for the $i$-th masked word. In a word, the training loss of the MLM task is the sum of the mean masked language modeling likelihood. The goal of the MDP task is to predict a set of real-valued descriptors of chemical characteristics, which is a regression task. The molecular descriptor encodes many physicochemical properties and can be easily calculated by RDKit \cite{landrum2006rdkit}. In this task, the final representation of the first token incorporates the global features of the entire molecule and is fed into a neural network model to predict the normalized set of descriptors. The task is optimized using the mean squared error over all predicted values as follows:

\begin{equation}
{\rm Loss}_{MDP} = \frac{1}{N_{desc}}\sum_{i=1}^{N_{desc}}(y_{i}-\hat{y}_{i})^2
\end{equation}
where $N_{desc}$ is the number of the molecular descriptors used in this task, $y_{i}$ is the normalized value of the $i$-th descriptors, and $\hat{y}_{i}$ is the predicted value of the $i$-th descriptors. The MFGP task can be formulated as a multi-label classification, which aims to predict the functional groups within the input molecule. The functional group contains rich domain knowledge of molecules and also can be easily detected by RDKit. The final representation of the first token is also fed into a neural network model for multi-label classification. This task is optimized using the cross-entropy loss as follows:

\begin{equation}
{\rm Loss}_{MFGP} = \frac{1}{N_{fun}}\sum_{i=1}^{N_{fun}}[y_i log(\hat{y_i}) + (1-y_i)(1-log(\hat{y_i}))]
\end{equation}
where $N_{fun}$ is the number of the molecular functional groups used in this task, $y_i$ is the true label indicating whether the molecule contains the $i$-th functional group, and $\hat{y_i}$ is the predicted probability of the $i$-th functional group. The final training loss of the self-supervised molecular representation learning model is given by the weighted sum of all individual task losses as follows:

\begin{equation}
{\rm Loss} = {\rm Loss}_{MLM} + \alpha{\rm Loss}_{MDP} + \beta{\rm Loss}_{MFGP}
\end{equation}
where $\alpha$ and $\beta$ are two weighting factors. The training objective is to minimize the loss and use backpropagation to optimize the model and update the representations. 

\subsubsection{Unsupervised protein representation learning}

In the target protein representation learning step, we employ ESM-2 \cite{Lin2022ESM}, a family of large-scale protein language models at scales from 8 million parameters up to 15 billion parameters, to extract the embeddings of target proteins. The ESM-2 language models also use the BERT-style \cite{devlin2018bert} encoder with transformer \cite{vaswani2017attention} architecture to train the masked language modeling objective, which aims to predict the original identity of randomly masked amino acids in a protein sequence based on their context. The UniRef \cite{suzek2015uniref} protein sequence database is used for the training of ESM-2 models, including $\sim$138 million UniRef90 sequences and $\sim$65 million unique sequences. 

The pre-trained ESM-2 models can directly predict the residue-residue contact map of the protein extracted from the Transformer self-attention patterns. Specifically, given a model with $L$ layers, $K$ heads, let $c_{ij}$ be a binary random variable, indicating whether the amino acids $i, j$ are in contact. Then the probability of contact between positions $i$ and $j$ is defined as a logistic regression:

\begin{equation}
p(c_{ij};\beta) = \frac{1}{1+{\rm exp}(-{\beta}_0-\sum_{l=1}^L\sum_{k=1}^{K}\beta_{kl}a_{ij}^{kl})}
\end{equation}
where $a_{ij}^{kl}$ is attention score between amino acids $i$ and $j$ from the $k$-th attention head in the $l$-th layer of the transformer.

And the ESM-2 language models are also enabled to generate high-resolution protein three-dimensional structure predictions from the protein sequence (ESMFold). In this work, we employ one of the ESM-2 models with 33 layers and 650 million parameters and use its hidden states of the last layer as the representations of target proteins.

\subsubsection{Downstream drug target prediction}

The last step is to integrate the drug and target representations and make various downstream predictions via AutoGluon \cite{erickson2020autogluon}. AutoGluon is an AutoML framework for structured data that automatically utilize state-of-the-art techniques without the need for frequent manual intervention to achieve strong predictive performance in many applications. Unlike prior AutoML frameworks that primarily focus on the task of Combined Algorithm Selection and Hyperparameter optimization (CASH) to find the best model from a sea of possibilities, AutoGluon performs advanced data processing and powerful multi-layer model ensembling to train highly accurate machine learning models. AutoGluon integrates various types of models (such as neural networks, LightGBM boosted trees and Random Forests), and ensembles these models based on novel combinations of multi-layer stacking and repeated $k$-fold bagging. 

In multi-layer stacking, the first layer has multiple base models, whose inputs are the original data features, and outputs are concatenated with data features and then fed into the next layer. And the last stacking layer leverages ensemble selection to aggregate the predictions of the stacker model in a weighted fashion. In the repeated $k$-fold bagging, the training data is randomly divided into $k$ disjoint chunks, each chunk is used as a test set to produce out-of-fold (OOF) predictions and the remaining chunks are used as a training set to train a model. To minor overfitting in OOF predictions, AutoGluon repeats the $k$-fold bagging process on $n$ different random partitions of the training data, and all OOF predictions are averaged over the repeated bags. More specifically, the training data ($X, Y$) is first randomly split into $k$ chunks $\{X_i^j, Y_i^j\}_{j=1}^k$ in the $i$-th repetition, then train a model on ($X_i^{-j}, Y_i^{-j}$) and make predictions $\hat{Y}_{m,i}^j$ on OOF data $X_i^j$ for each model type $m$ in the family of models $\mathbf{M}$. The outputs of model type $m$ in the stacking layer $l$ are computed by averaging all OOF predictions over the repeated bags, that is,
\begin{equation}
\hat{Y}_{m} = \{\frac{1}{n} {\sum}_i \hat{Y}_{m,i}^j\}_{j=1}^k
\end{equation}
which are concatenated with the inputs and then fed into the next stacking layer, that is,
\begin{equation}
X \leftarrow {\rm concatenate}(X, \{\hat{Y}_{m}\}_{m \in \mathbf{M}})
\end{equation}
The final predictions are the aggregation of the stacker models’ predictions in a weighted manner.

The framework, in this work, is highly adaptable and can be utilized for various drug target prediction tasks, including DTI, DTA, and MoA. This framework employs pre-trained drug and target representation learning models that can be shared across different tasks, and are then fine-tuned using distinct labeled datasets in a supervised learning manner. The pre-training phase enables the models to learn precise representations from a vast amount of unlabeled data, leading to an exceptional performance on downstream tasks.

\subsection{Benchmark datasets} 

In this study, six benchmark datasets for three prediction tasks (i.e., DTI, DTA, and MoA), namely Activation, Inhibition, Yamanishi\_08, Hetionet, Davis, and KIBA, were used to comprehensively evaluate the performance and ability of our model. 

Activation and Inhibition are two MoA datasets that were obtained from the Therapeutic Target Database (TTD) \cite{zhou2022therapeutic}. We selected those MOAs that are explicitly defined as activation (e.g., ``activator'', ``agonist'') or inhibition (e.g., ``inhibitor'', ``antagonist''). In total, we obtained 1913 activation MoAs between 1426 drugs and 281 targets for the Activation dataset, and 21055 inhibition MoAs between 14049 drugs and 1088 targets for the Inhibition dataset.

The Yamanishi\_08 dataset and Hetionet are DTI datasets, in which the labels are binary interactions between drugs and targets. The Yamanishi\_08’s dataset is originally introduced by Yamanishi et al. \cite{yamanishi2008prediction} and consists of four sub-datasets: G-Protein Coupled Receptors (GPCR), Ion Channels (IC), Nuclear Receptors (NR) and Enzymes (E) obtained from KEGG BRITE, BRENDA, SuperTarget, and DrugBank databases \cite{kanehisa2006genomics,schomburg2004brenda,gunther2007supertarget,wishart2008drugbank}. In this study, we use the combined dataset of the four sub-datasets constructed by Ye et al. \cite{ye2021unified}. In total, the dataset contains 791 drugs, 989 targets, and 5127 known DTIs (i.e. positive samples). The Hetionet dataset is constructed by Himmelstein et al. \cite{himmelstein2017systematic}, which integrated the biomedical data from 29 public resources. It comprises 1384 drugs, 5763 targets, and 49942 DTIs.

Davis and KIBA are two DTA datasets and are popular standard datasets used in previous work for DTA prediction evaluation \cite{ozturk2018deepdta,he2017simboost}. The Davis dataset contains binding intensities of the kinase protein family and the relevant inhibitors measured using dissociation constant (${\rm K_d}$) values. It consists of 68 drugs and 442 targets and was constructed by Davis et al. \cite{davis2011comprehensive}. KIBA is a large-scale kinase inhibitor bioactivity dataset constructed by Tang et al. \cite{tang2014making}. It combined different measurement types such as ${\rm K_i}$, ${\rm K_d}$ and ${\rm IC_{50}}$, and contains 2111 drugs and 229 targets.

\subsection{Baselines} 

In this work, three types of baseline models are used in the performance comparison for the DTI, DTA, and MoA prediction tasks, including CPI\_GNN \cite{tsubaki2019compound}, TransformerCPI \cite{chen2020transformercpi}, MPNN\_CNN \cite{gilmer2017neural}, and KGE\_NFM \cite{ye2021unified} for DTI prediction, and DeepDTA \cite{ozturk2018deepdta}, MONN \cite{li2020monn}, BACPI \cite{li2022bacpi}, and GraphDTA \cite{nguyen2021graphdta} for DTA prediction, and AI-DTI \cite{lee2021predicting} for MoA prediction. CPI\_GNN, TransformerCPI, MPNN\_CNN, DeepDTA, BACPI, and GraphDTA as well as DTIAM require only SMILES strings of compounds and primary sequences of proteins as input. KGE\_NFM requires the heterogeneous information extracted from multi-omics data to build a knowledge graph and used the Morgan fingerprints of drugs and CTD descriptors of targets as additional information. MONN requires not only SMILES strings and protein sequences, but also pairwise non-covalent interactions between atoms of compounds and residues of proteins as extra supervision information. In this work, since the non-covalent interactions labels of Davis and KIBA datasets were unavailable, we did not provide the extra supervision information for MONN (denoted as ${\rm MONN_{single}}$ in the original paper). AI-DTI needs SMILES strings of compounds and genetically perturbed transcriptome data of target genes as input. Note that, the MPNN\_CNN and DeepDTA models were constructed with DeepPurpose \cite{huang2020deeppurpose}.

\subsection{Experimental settings} 

DTIAM and these baseline methods are evaluated under three different settings of cross-validation, i.e., the warm start setting, the drug cold start setting and the target cold start setting. To explain these settings, we denote the training and test drug sets by $D_{train}$ and $D_{test}$, and training and test drug sets by $T_{train}$ and $T_{test}$, respectively, and use ($d_i$, $t_j$) to represent the drug-target pair between the $i$-th drug and $j$-th target.

In the warm start setting, for a drug-target pair ($d_i$, $t_j$) from the test set ($d_i \in D_{test}$ and $t_j \in T_{test}$), both $d_i$ and $t_j$ are encountered in the training set ($d_i \in D_{train}$ and $t_j \in T_{train}$). That is, the test and training sets share common drugs and targets. This scenario is suitable for identifying potential interactions between known drugs and targets.

In the drug cold start setting, for a drug-target pair ($d_i$, $t_j$) from the test set, the drug $d_i$ is
unseen in the training phase ($d_i \in D_{test}$, $d_i \notin D_{train}$), while the target $t_j$ is present in both training and test sets ($t_j \in T_{test}$, $t_j \in T_{train}$). This experimental setting is relevant if we need to discover potential candidate targets for new drugs.

In the target cold start setting, for a drug-target pair ($d_i$, $t_j$) from the test set, we have seen the drug $d_i$ ($d_i \in D_{train}$), but the target $t_j$ is unseen in the training phase ($t_j \notin T_{train}$). This scenario is often applied in virtual drug screening of new targets.

Note that the DTI prediction task is evaluated under 10-fold cross-validation, and the DTA and MoA prediction tasks are evaluated under 5-fold cross-validation. In addition, for the KGE\_NFM model, the train-test split schemes of the drug/target cold start setting only focus on the drugs/proteins existing in the knowledge graph but without any known DTI relations. 

\subsection{Evaluation metrics} 

In this study, we use the average scores of the area under the receiver operating characteristics curve (AUROC) and the area under the precision-recall curve (AUPR) to evaluate the performance of each method in the DTI and MoA prediction tasks. In this work, we set the ratio between the positive and negative samples to 1:10 because this is more in line with real-world scenarios. Here, we adopt AUPR as the main evaluation metric, since it gives a more accurate evaluation of a method’s performance in the unbalanced dataset. In the DTA prediction task, the performance of each method was evaluated by the mean squared error (MSE) and the Pearson correlation. 

\backmatter

\section*{Acknowledgments}

This work was supported by the National Natural Science Foundation of China [61832019], the Hunan Provincial Science and Technology Program [2019CB1007], the Science and Technology Innovation Program of Hunan Province [2021RC4008]. We acknowledge the High Performance Computing Center of Central South University for support.

\emph{Conflict of Interest}: none declared.

\section*{Availability of data and code}

The source data and codes of DTIAM are available on GitHub at https://github.com/CSUBioGroup/DTIAM

\bibliography{ref}

\end{document}